\documentclass[aps,prl,twocolumn,showpacs,floatfix]{revtex4}

\usepackage{amsmath}
\usepackage{graphicx}
\usepackage{dcolumn}
\usepackage{bm}

\newcommand{\kb}{k_{\rm B}}

\begin{document}
\title{The Transition State in a Noisy Environment}
\author{Thomas Bartsch}
\author{Rigoberto Hernandez}
\author{T. Uzer}
\affiliation{Center for Nonlinear Science,
  Georgia Institute of Technology, Atlanta, GA 30332-0430, USA}
\date{\today}

\begin{abstract}
Transition State Theory overestimates reaction rates in solution because
conventional dividing surfaces between reagents and products are crossed
many times by the same reactive trajectory.  We describe a recipe for
constructing a time-dependent dividing surface free of such recrossings in
the presence of noise.  The no-recrossing limit of Transition State Theory
thus becomes generally available for the description of reactions in a
fluctuating environment.
\end{abstract}

\pacs{82.20.Db, 34.10.+x, 05.40.Ca, 05.45.-a}
\maketitle

Chemical reaction rates are determined by what happens in a very small
fraction of phase space: Such is the premise of Transition State Theory
(TST). It revolves around the concept of the Transition State (TS), also
known as the activated complex, that is defined by the dividing surface
that best separates ``reactant'' and ``product'' regions in phase space.
Actually, the TS is a general structure found in all dynamical systems that
evolve from ``reactants'' to ``products''.  In addition to chemical
reaction dynamics \cite{Miller}, it also determines rates in many other
interesting systems such as asteroid capture \cite{genesis}, mass transport
through the solar system \cite{Koon00}, the rearrangements of clusters
\cite{Komatsuzaki02}, the ionization of atoms \cite{JFU00}, conductance due
to ballistic electron transport through microjunctions \cite{ballistic},
and diffusion jumps in solids \cite{toller}.

TST is central to the understanding of chemical reactions because its
radical simplification captures the physics involved in a pictorial way,
but its accuracy hinges on the so-called ``no-recrossing'' requirement: If
reagents and products are separated by an abstract hypersurface in phase
space, the rate will be exact if each reacting trajectory crosses this
surface once and only once. Recrossings will lead to an overestimate of the
rate, a problem that occupied Wigner in the early days of TST
\cite{wigner38,Wigner39}.  While in low-dimensional systems a solution has
been known for a long time \cite{pech73}, enforcing the no-recrossing
condition in more than two degrees of freedom has been an open problem
until very recently \cite{Komatsuzaki99,bristol1,bristol2}. It is even more
difficult to uphold for reactions in randomly fluctuating environments
(such as chemical and biological processes in liquids) where the noise
causes the particle to move randomly back and forth, so that it will
typically cross any fixed dividing surface in phase space many times
\cite{talkner95}.  To solve this long-standing problem, many approximate TS
structures have been suggested in the literature
\cite{Keck67,hynes85a,Truhlar96,pollak86b,Tucker95,Pollak00,Martens02}.
Martens \cite{Martens02}, in particular, constructed a stationary
stochastic separatrix (the collection of all phase-space points for which
the reaction probability is 1/2) that foreshadows the time-dependent
structures to be described here. In some cases, infinite dimensional
representations of the Langevin equation
\cite{zwan73,caldeira81,pollak86b}, in which the system is coupled to a
bath of harmonic oscillators, have been used to obtain approximate
infinite-dimensional dividing surfaces that lead to excellent
approximations to the rate \cite{pollak86b,pgh89,Graham90}.

In this Letter, an exact transition state in a noisy environment 
is constructed as a dividing surface which moves so as to avoid recrossings.  
For each instance of the noise, there is a unique trajectory which
remains in the vicinity of the barrier for all time. This trajectory serves
as the origin of a moving coordinate system to which the geometric
structures of the noiseless phase space are attached.  The result is a
moving dividing surface, free of recrossings, that describes a reaction
influenced by noise in the same way as a static dividing surface does in
conventional TST.  
This construction goes beyond the
geometric approach of~\cite{Martens02} by fully taking into account the
time-dependence of the fluctuating force. Indeed, the moving separatrices
introduced here determine with certainty if a trajectory will or will not
react.

We illustrate our findings on a reactive system described by the Langevin
equation \cite{zwan01}
\begin{equation}
  \label{LE}
  \ddot{\vec q}_\alpha(t) = -\nabla_{\vec q} U(\vec q_\alpha(t)) -
        {\bm\Gamma}\dot{\vec q}_\alpha(t) +
        \vec\xi_\alpha(t) \;.
\end{equation}
The vector $\vec q$ denotes a set of $N$ mass-weighted coordinates,
$U(\vec q)$ the potential of mean force governing the reaction, $\bm\Gamma$
a symmetric positive-definite friction matrix and $\vec\xi_\alpha(t)$ a
fluctuating force assumed to be Gaussian with zero mean.  The subscript
$\alpha$ represents randomness by labeling different instances of the
fluctuating force. The latter is related to the friction matrix $\bm\Gamma$
by the fluctuation-dissipation theorem \cite{zwan01}
\begin{equation}
  \label{FDTWhite}
  \left<\vec\xi_\alpha(t)\vec\xi_\alpha^\text{T}(t')\right>_\alpha = 
      2\kb T\, {\bm\Gamma} \, \delta(t-t') \;,
\end{equation}
where the angular brackets denote the average over the instances $\alpha$
of the noise.

The reactant and product regions in configuration space are separated by a
potential barrier whose position is marked by a saddle point $\vec
q_0^\ddag=0$ of the potential $U(\vec q)$. The reaction rate is primarily
determined by the dynamics in a small neighborhood of $\vec q_0^\ddag$
\cite{Truhlar96,Miller,hynes85a,pollak86b,Tucker95,Pollak00}, so that 
the deterministic force can be linearized around the saddle point to obtain
\begin{equation}
  \label{linLE}
  \ddot{\vec q}_\alpha(t) = {\bm\Omega} \vec q_\alpha(t) -
        {\bm\Gamma}\dot{\vec q}_\alpha(t) + \vec\xi_\alpha(t)
\end{equation}
with a symmetric matrix $\bm\Omega$ given by $\Omega_{ij}=-\left(\partial^2
U/\partial q_i \partial q_j\right)_{\vec q=\vec q_0^\ddag}$.  
Coordinates can always be chosen to that $\bm{\Omega}$ is diagonal.
It must have
one positive eigenvalue $\omega_\text{b}^2$ equal to the squared barrier
frequency and $N-1$ negative eigenvalues $-\omega_i^2$, where $\omega_i$
are the frequencies of transverse oscillations.  Including
velocity-dependent forces from, e.g., a magnetic field \cite{JFU00} in this
setup would be straightforward.  Note that no restriction on the
dimensionality of the configuration space is imposed.

Our aim in the following is to construct a recross\-ing-free surface for
equation~(\ref{linLE}).  The difference
\begin{equation}
  \label{relCoord}
  \Delta\vec q(t)=\vec q_\alpha(t)-\vec q_\alpha^\ddag(t)
\end{equation}
of two trajectories under the influence of the same fluctuating force,
which describes the location of the trajectory $\vec q_\alpha(t)$ relative
to the moving origin $\vec q_\alpha^\ddag(t)$, satisfies a deterministic
equation of motion.  Since this difference is independent of noise, there
is no need to refer to $\alpha$ in $\Delta\vec q(t)$.  Thus,
invariant manifolds and a no-recrossing surface can be specified for the
relative dynamics. These geometric objects in the noiseless phase space can
then be regarded as being attached to the reference trajectory $\vec
q_\alpha^\ddag(t)$ while being carried around by it.  They define randomly
moving invariant manifolds and a moving no-recrossing surface in the phase
space of the original, noisy system.

The construction outlined above can be carried out for any reference
trajectory $\vec q_{\alpha}^\ddag(t)$ and leads to a multitude of
no-recrossing surfaces, but only one specific surface is relevant to the
reaction dynamics: The crossing of this surface should signal the transition
of the trajectory $\vec q_\alpha(t)$ from the reactant side to the product
side of the barrier.  If the reference trajectory is chosen arbitrarily, it
will typically descend into either the reactant or product wells over
time. However, only a reference trajectory that remains in the vicinity of
the barrier can carry a no-recrossing surface that actually describes the
reaction.  Indeed, as shown here, for each instance of the noise there
is a unique reference trajectory with this property. It represents, in
mathematical terms, an invariant measure of the noisy dynamical system
\cite{Arnold98}. We call it the Transition State Trajectory and henceforth
restrict the notation $\vec q_\alpha^\ddag(t)$ to this privileged reference
trajectory.

We solve equation~(\ref{linLE}) by rewriting it as a first-order equation
in $2N$-dimensional phase space with coordinates $\vec x=(\vec q, \vec v)$,
with $\vec v=\dot{\vec q}$, and diagonalizing its linear part.
Equation~(\ref{linLE}) then decomposes into a set of $2N$ independent
scalar equations
\begin{equation}
  \label{diagEq}
  \dot x_{\alpha j}(t) = \epsilon_j x_{\alpha j}(t) + \xi_{\alpha j}(t) \;,
\end{equation}
where $\epsilon_j$ are the eigenvalues of the linear part, $x_{\alpha j}$
the components of $\vec x$ in a basis of eigenvectors and $\xi_{\alpha j}$
the corresponding components of $(0,\vec\xi_\alpha(t))$.

\begin{figure}
  \centerline{\includegraphics[width=.44\textwidth]{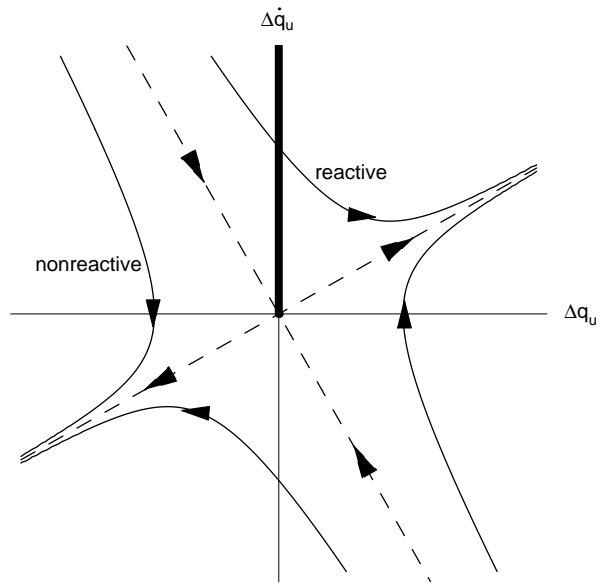}}
  \caption{\label{fig:phase}
     Phase portrait of the one-dimensional noiseless damped dynamics
     corresponding to equation~(\ref{linLE}). The thick
     line indicates a possible choice for the no-recrossing surface. The
     simple relative dynamics illustrated here is exemplified by the
     trajectories in figure~\ref{fig:Traj2D}(a).
  }
\end{figure}

The components $\Delta q_j(t)$ of the relative coordinate~(\ref{relCoord})
satisfy a noiseless version of~(\ref{diagEq}) with $\xi_{\alpha j}=0$. They
grow or decay exponentially, depending on whether the eigenvalue
$\epsilon_j$ has positive or negative real part.  In $N=1$ degree of
freedom there is one positive and one negative real eigenvalue.  The phase
portrait of the dynamics is shown in figure~\ref{fig:phase}. The
eigenvectors span one-dimensional stable and unstable manifolds of the
saddle point. They act as separatrices between reactive and non-reactive
trajectories. The knowledge of the invariant manifolds allows one to
determine the ultimate fate of a specific trajectory from its initial
conditions. It is also easy to identify lines in the quadrant of reactive
trajectories that are surfaces of no recrossing. Clearly, the half-line
$\Delta q=0$, with $\Delta \dot q>0$ serves this purpose.

\begin{figure*}
  \centerline{\includegraphics[width=.29\textwidth]{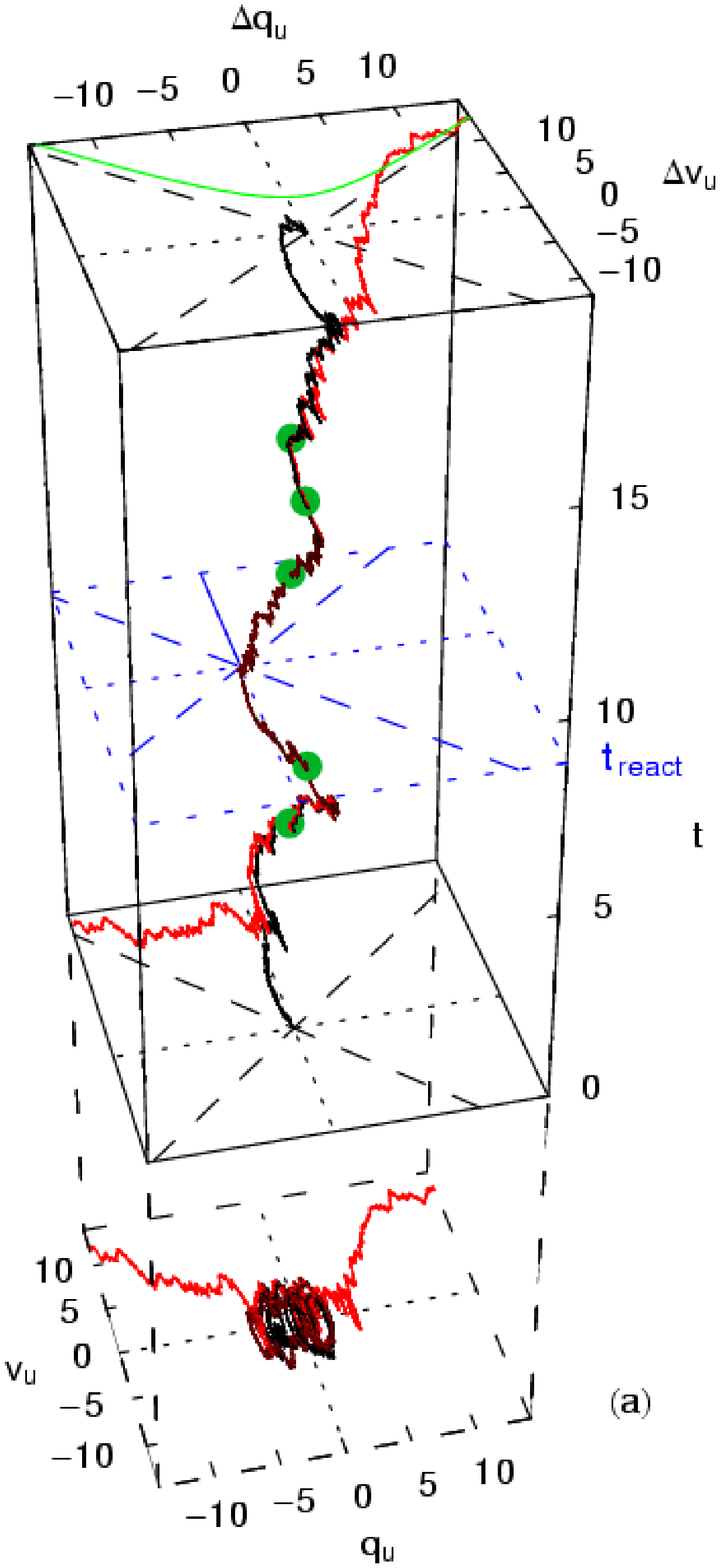}\hfill
              \includegraphics[width=.29\textwidth]{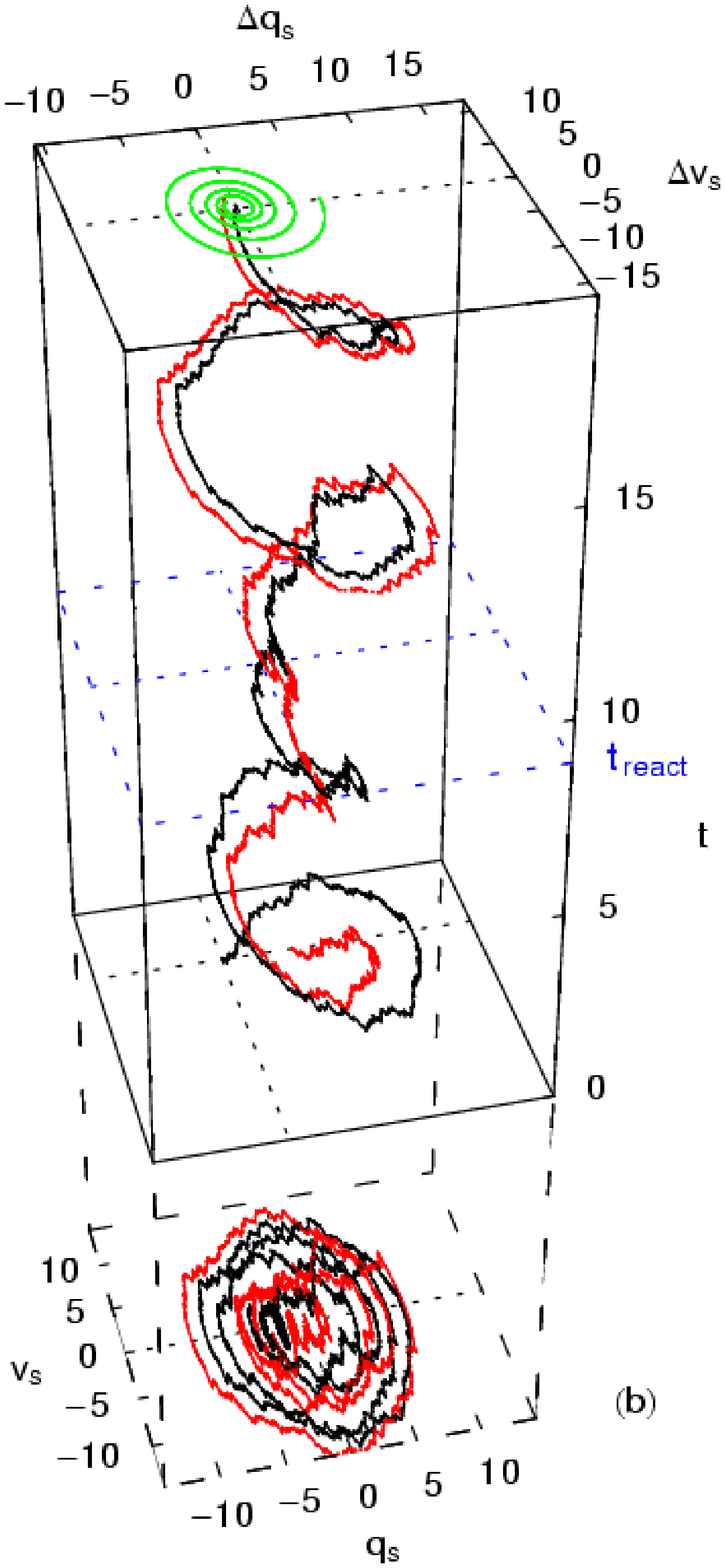}\hfill
              \includegraphics[width=.29\textwidth]{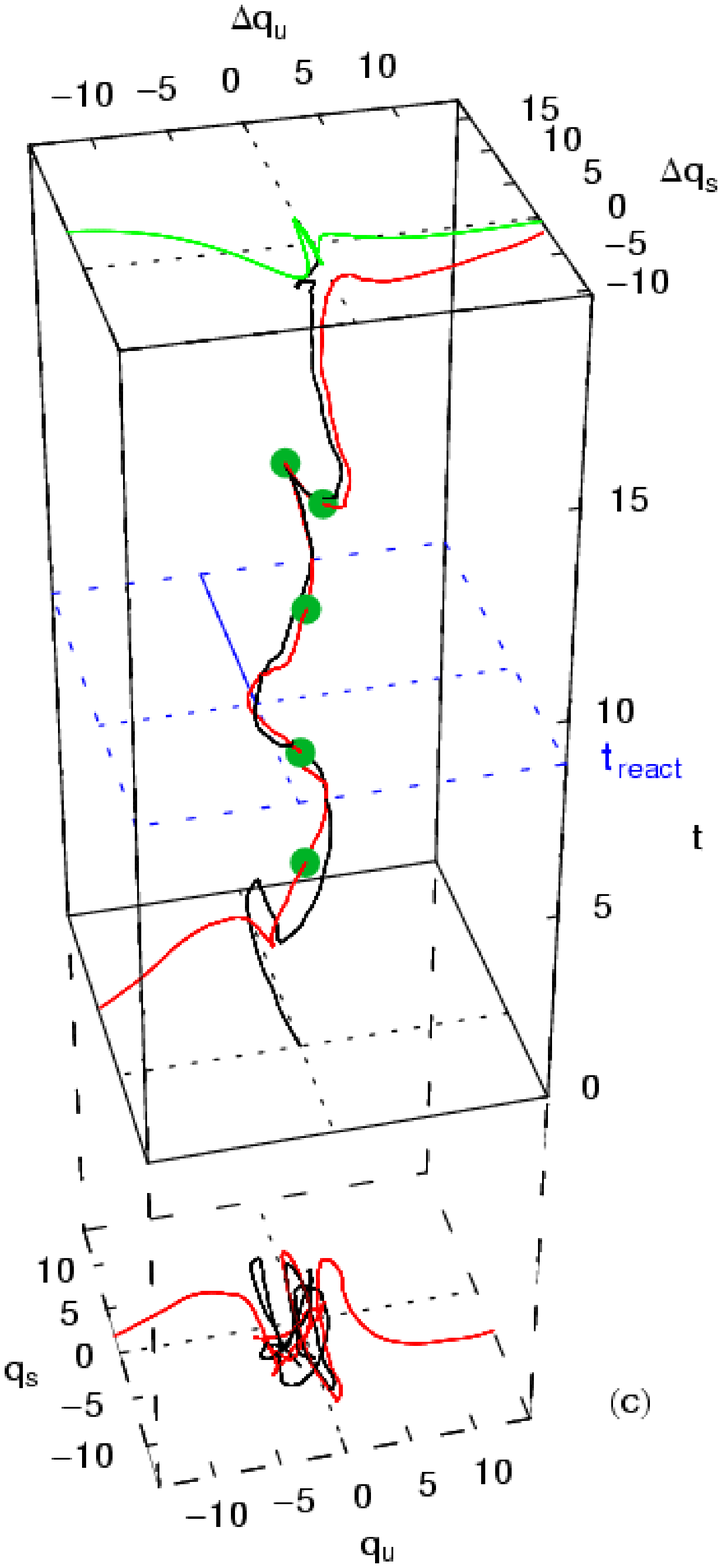}}
  \caption{\label{fig:Traj2D}
    A random instance of the TS trajectory (black) and a reactive
    trajectory (red) under the influence of the same noise in a system with
    $N=2$ degrees of freedom:
the unstable reactive coordinate $q_\text{u}$ 
and the stable transverse coordinate $q_\text{s}$.
The potential is
$U(q_\text{u}, q_\text{s}) = -\frac 12 \omega_\text{b}^2 q_\text{u}^2
                               +\frac 12 \omega_\text{s}^2 q_\text{s}^2$.
    Trajectories
    are projected onto (a) the reactive degree of freedom, (b) the
    transverse degree of freedom, and (c) configuration space. Units are
    chosen so that $\omega_\text{b}=1$ and $\kb T=1$.  The transverse
    frequency is $\omega_\text{s}=1.5$, and the friction is isotropic,
    $\bm\Gamma=\gamma\mbox{\bf I}$, with $\gamma=0.2$.  The bottom of each
    column shows the projected trajectories in the corresponding space.
    Above this, their time
    evolution is illustrated using the same axes. The blue cut 
    marks the unique reaction time $t_\text{react}=8.936$,
    when the moving TS surface is
    crossed. Dotted lines in this cut, at $t=0$, and at $t=20$ indicate 
    the moving coordinate axes centered on the TS
    Trajectory.  These axes are labeled explicitly only at the top face.
    Dashed lines in the cuts of column (a)
    show the moving invariant manifolds.
    No TS is indicated in column (b) because the
    $q_\text{s}-v_\text{s}$ subspace lies entirely within the moving
    TS surface. Thick green dots indicate repeated crossings of the
    stationary TS surface $q_\text{u}=0$.
    Green lines in the top face of all three columns show
    the reactive trajectory in relative coordinates $\Delta\vec
    q$, $\Delta\vec v$ (not to scale in (a), for graphical
    reasons). The typical behavior of a reactive trajectory shown in
    figure~\ref{fig:phase} is visible in the top face of 
    column (a).
}
\end{figure*}

In multiple degrees of freedom, transverse damped oscillations must be
added to the phase portrait in figure~\ref{fig:phase}. Their presence
manifests itself through $N-1$ complex conjugate pairs of eigenvalues
$\epsilon_j$. For strong damping, some of the transverse modes can become
overdamped, so that further eigenvalues become negative real. In any case,
there is exactly one positive real eigenvalue which corresponds to the
particle sliding down the barrier. In all other directions, the dynamics is
stable. The eigenvector corresponding to the smallest negative eigenvalue
together with the unstable eigenvector span a plane in phase space in which
the dynamics is given by the phase portrait of figure~\ref{fig:phase}. The
separatrices between reactive and nonreactive trajectories that were
identified for the one-dimensional dynamics together with the stable
transverse subspace form separatrices in the high-dimensional phase
space. In a similar manner, a no-recrossing surface in the full phase space
is spanned by a no-recrossing curve in the plane and the transverse
directions.

Using the Green function technique and assuming $\Re\epsilon_j<0$, we find
a particular solution
\begin{equation}
  \label{xiStab}
  \begin{split}
    x^{{\rm s}\,\ddag}_{\alpha j}(t)
        &= \int_{-\infty}^0 e^{-\epsilon_j \tau} \xi_{\alpha j}(t+\tau)\,d\tau
  \end{split}
\end{equation}
of equation~(\ref{diagEq}). The general solution consists of~(\ref{xiStab})
plus an exponential term that must be suppressed to keep the solution
bounded as $t\to-\infty$. Similarly, for unstable components with
$\Re\epsilon_j>0$, the only solution to~(\ref{diagEq}) that remains bounded
for $t\to+\infty$ is
\begin{equation}
  \label{xiUnst}
  \begin{split}
    x^{{\rm u}\,\ddag}_{\alpha j}(t) 
      &= -\int_0^\infty e^{-\epsilon_j\tau} \xi_{\alpha j}(t+\tau) \, d\tau \;.
  \end{split}
\end{equation}
Equations~(\ref{xiStab}) and~(\ref{xiUnst}) specify all the phase space
components of the unique random trajectory ---{\it viz.} the TS Trajectory---
that remain bounded in the remote future as well as in the distant past.
Its configuration space representation is obtained by
transforming back to position and velocity coordinates. Because the
components $\xi_{\alpha j}$ of the fluctuating force are Gaussian random
variables with zero mean, so are the components~(\ref{xiStab})
and~(\ref{xiUnst}) of the TS trajectory. 
Their joint probability
distribution can readily be specified through the explicit
calculation of their cross-correlation function
by insertion of the correlation function~(\ref{FDTWhite}) of the noise.

As described above, the TS trajectory serves as the origin of a moving
coordinate system to which the geometric structures in the noiseless phase
space are attached. This is illustrated in figure~\ref{fig:Traj2D} for
$N=2$ degrees of freedom. 
The phase space is
four-dimensional and already exhibits all the salient features of the
dynamics in arbitrarily high dimensions. The figure compares a TS
trajectory with a reactive trajectory under the influence of the same
noise. The trajectory approaches the TS trajectory from the reactant side
$q_\text{u}<0$, remains in its vicinity for a while and then wanders off to
the product side. Because the moving invariant manifolds of the TS
trajectory are known, it can be predicted already at time $t=0$ that the
trajectory actually will lead to a reaction instead of returning to the
reagent side of the saddle.

The reactive trajectory crosses the space-fixed TS surface $q_\text{u}=0$
several times before it finally descends on the product side. The moving TS
surface $\Delta q_\text{u}=0$, by contrast, is crossed only once, at the
reaction time $t_\text{react}$ indicated by blue lines. That this is
actually the case can be seen from the curves in the top faces of each
column, which indicate the noiseless relative motion $\Delta\vec q(t)$,
$\Delta\vec v(t)$ between the TS trajectory and the reactive
trajectory. The relative coordinates show, as expected, the hyperbolic
motion known from figure~\ref{fig:phase} in the unstable degree of freedom,
a damped oscillation in the stable transverse degree of freedom, and a
superposition of the two in the configuration-space projection in
figure~\ref{fig:Traj2D}(c).

The derivations presented here are based on the assumption~(\ref{FDTWhite})
of white noise.  If the noise were correlated, the friction would contain
memory and the dynamics would consequently be significantly more complicated
\cite{zwan01}. The construction of the TS
trajectory can be carried out nonetheless.  It leads to
integral formulas similar to~(\ref{xiStab}) and~(\ref{xiUnst}), with the
eigenvalues $\epsilon_j$ given by the Grote-Hynes equation
\cite{grot80,pollak86b}. These generalizations will be presented in a
forthcoming publication.

In recent simulations \cite{chan98a,chan02a}, ensembles of transition
paths, i.e., trajectories connecting reactants to products, have been shown
to be useful in computing rates even for high-dimensional systems.  Our
identification of the TS trajectory provides additional insight into the
geometry of reaction dynamics that complements the transition path ensemble
method. While the TS trajectory introduced here is not itself a transition
path, it carries a dividing surface that is transverse to all transition
paths and is crossed once and only once by each of them.  It is no harder
to compute than a typical trajectory.  This is to be contrasted with most
TST approaches in which finding a no-recrossing dividing surface requires
the onerous solution of a highly nonlinear problem.

In summary, to generalize the formalism of TST to reactive systems driven
by noise, a moving surface has been constructed that is crossed once and only
once by each transition path. This surface is therefore suited to take over
the role of the well-known and widely applied static dividing surface of
conventional TST in a broader time-dependent setting.

\begin{acknowledgments}
This work was partly supported by the US National
Science Foundation and by the Alexander von Humboldt-Foundation.
\end{acknowledgments}


\end{document}